# Unsupervised Domain Adaptation for Automated Knee Osteoarthritis Phenotype Classification

Junru Zhong[1,2]*, **MS**, Yongcheng Yao[1,2]*, **MPhil**, Dónal G. Cahill[1,2], **FFR RCSI**, Fan Xiao[3], **PhD**, Siyue Li[1,2], **MS**, Jack Lee[4], **PhD**, Kevin Ki-Wai Ho[5], **FRCS Ortho (Edin)**, Michael Tim-Yun Ong[5], **FRCS Ortho (Edin),** James F. Griffith[1,2], **MD**, Weitian Chen[1,2], **PhD**

*Authors share the same contribution.*

[1]CU Lab for AI in Radiology (CLAIR), Department of Imaging and Interventional Radiology, The Chinese University of Hong Kong

[2] Department of Imaging and Interventional Radiology, The Chinese University of Hong Kong

[3] Department of Radiology, Shanghai Sixth People's Hospital Affiliated to Shanghai Jiao Tong University School of Medicine

[4] Centre for Clinical Research and Biostatistics, The Chinese University of Hong Kong

[5] Department of Orthopaedics & Traumatology, The Chinese University of Hong Kong

# Abstract

**Purpose**: The aim of this study was to demonstrate the utility of unsupervised domain adaptation (UDA) in automated knee osteoarthritis (OA) phenotype classification using a small dataset ($n = 50$).

**Materials and Methods**: For this retrospective study, we collected 3,166 three-dimensional (3D) double-echo steady-state magnetic resonance (MR) images from the Osteoarthritis Initiative dataset and 50 3D turbo/fast spin-echo MR images from our institute (in 2020 and 2021) as the source and target datasets, respectively. For each patient, the degree of knee OA was initially graded according to the MRI Osteoarthritis Knee Score (MOAKS) before being converted to binary OA phenotype labels. The proposed UDA pipeline included (a) pre-processing, which involved automatic segmentation and region-of-interest cropping; (b) source classifier training, which involved pre-training phenotype classifiers on the source dataset; (c) target encoder adaptation, which involved unsupervised adaption of the source encoder to the target encoder and (d) target classifier validation, which involved statistical analysis of the target classification performance evaluated by the area under the receiver operating characteristic curve (AUROC), sensitivity, specificity and accuracy. Additionally, a classifier was trained without UDA for comparison.

**Results**: The target classifier trained with UDA achieved improved AUROC, sensitivity, specificity and accuracy for both knee OA phenotypes compared with the classifier trained without UDA.

**Conclusion**: The proposed UDA approach improves the performance of automated knee OA phenotype classification for small target datasets by utilising a large, high-quality source dataset for training. The results successfully demonstrated the advantages of the UDA approach in classification on small datasets.

# Introduction

Osteoarthritis (OA) is a common degenerative disease. Ageing populations worldwide contribute to the increasing demand for OA diagnosis, staging and grading (1). Manual OA grading of magnetic resonance (MR) images using semi-quantitative systems, like the MRI Osteoarthritis Knee Score (MOAKS) (2), are time-consuming and subjective. Advances in deep learning have enabled the development of techniques for automatically grading the severity of knee cartilage damage (3), anterior cruciate ligament injury (4,5) and OA in radiography (6) and MR (7) images. However, these deep learning-based methods require large labelled datasets.

Unsupervised domain adaptation (UDA) is a transfer learning technique that enables convolutional neural networks (CNN) trained on a source domain to be adapted to a target domain without ground truth labels from the target data (8). UDA is one of the primary methods for transferring information from larger and more generalised datasets ('source data') to smaller datasets ('target data'). The UDA process enables downstream research on target data with zero labels (9). In medical image analysis, UDA allows researchers to take advantage of high-cost medical image datasets to expedite local research without additional labelling costs. UDA approaches align the source and target domains using various methods, such as domain translation (10), statistical matching and adversarial learning (11). Jiang et al. (8) and Guan and Liu (12) have published comprehensive reviews of UDA and its application in medical image analysis.

The application of UDA for OA grading or phenotype classification has not been explored. In this study, we proposed a novel application of UDA based on the Adversarial Discriminative Domain Adaptation (ADDA) (11) framework. We used a CNN trained on a publicly available dataset for automated knee OA phenotype classification on a small dataset ($n = 50$) obtained from our hospital. We implemented a systematic UDA approach for automatic phenotype classification. In this study, we (i) proposed a UDA approach for automatic and objective MRI-based OA phenotype classification to address the challenges of collecting a large labelled dataset associated with supervised techniques; (ii) compared the proposed method with a classifier trained without UDA, demonstrating the improved grading performance of our UDA approach and (iii) explored the application of UDA for OA phenotype classification tasks using datasets from multiple MRI vendors, acquisition protocols and research institutes.

## Material and Methods

We conducted a retrospective study to demonstrate the feasibility of UDA for MRI-based OA phenotype classification.

### Data and MRI Acquisition

The source dataset was a subset of the Osteoarthritis Initiative (OAI) dataset (13) comprised baseline and 4-year follow-up MRI scans and corresponding MOAKS scores. The MOAKS grades were graded on independently selected OAI subsets by experienced radiologists and have been previously reported (14–17). The detailed grading scheme is available on the OAI website (https://nda.nih.gov/oai). We combined the MOAKS sub-grades from these projects to form the source dataset and selected knee subjects with MR images and the required MOAKS sub-grades. The MR images were collected using a three-dimensional (3D) double-echo steady-state sequence according to the protocol reported by Peterfy et al. (13). We constructed a balanced source dataset by randomly selecting samples from the source dataset, keeping one-third of the samples in the balanced source dataset are patients.

The target dataset contained MRI scans of 50 knee subjects collected at our institute in 2020 and 2021. The Institutional Review Board approved the project, and all participants gave informed consent. Forty patients with radiographic OA (Kellgren–Lawrence grades 1–4, 10 patients per grade) and 10 healthy controls received knee MRI exams. Our target datasets were collected using three two-dimensional (2D) turbo/fast spin echo (TSE/FSE) MRI sequences and a 3D TSE/FSE sequence (VISTA™) on a Philips Achieva TX 3.0T scanner (Philips Healthcare, Best, Netherlands). The 2D scans were used for manual MOAKS grading, and the 3D VISTA™ scans were used for automated OA phenotype classification using the UDA pipeline. The detailed MRI protocol is reported in Table 1.

Table 1. Magnetic resonance imaging protocol of the target dataset

| Scan | 3D VIEW PD TSE | PD aTSE SAG | T2 SPAIR TSE COR | PD SPAIR TSE AX |
|---|---|---|---|---|
| Plane | Sagittal | Sagittal | Coronal | Axial |
| Fat Suppression | SPAIR | None | SPAIR | SPAIR |
| No. of slices | 150 | 25 | 25 | 25 |
| Field of View (mm$^3$) | $160 \times 160 \times 120$ | $162 \times 160 \times 82$ | $160 \times 160 \times 82$ | $150 \times 150 \times 82$ |
| TE/TR (ms/ms) | 26/900 | 30/3451 | 62/5429 | 30/5864 |
| X-resolution (mm) | 0.71 | 0.4 | 0.227 | 0.293 |
| Y-resolution (mm) | 0.71 | 0.546 | 0.227 | 0.293 |
| Scan time (min: sec) | 05:51 | 03:24 | 03:43 | 03:37 |
| Usage | Deep learning | MOAKS | MOAKS | MOAKS |

Note. The 3D VIEW PD TSE sequence was used for automatic phenotype classification. The remaining three sequences were prepared for MRI Osteoarthritis Knee Score (MOAKS) grading. 3D = three-dimensional, PD = proton density, TSE = turbo spin echo, SAG = sagittal, SPAIR = Spectral Attenuated Inversion Recovery, COR = coronal, AX = axial, TE = echo time, TR = repetition time, MOAKS = MRI Osteoarthritis Knee Score

## MOAKS Grades and Phenotype Labels

The MOAKS grades of the source dataset were obtained from the OAI, and the MOAKS grades for the target dataset were independently graded by two musculoskeletal radiologists following the grading protocol used by the OAI. Both radiologists had more than 6 years of experience. The MOAKS grades prepared by the two radiologists had an excellent intraclass correlation coefficient of 0.999 ($p < 0.01$).

The MOAKS grades were converted to binary phenotype labels using a Python script, according to the definitions recommended by Roemer et al. (18). The cartilage/meniscus phenotype was defined by meniscus damage and medial and lateral cartilage damage. The subchondral bone phenotype was defined by bone marrow lesions on the tibiofemoral joint (TFJ) and patellofemoral joint (PFJ).

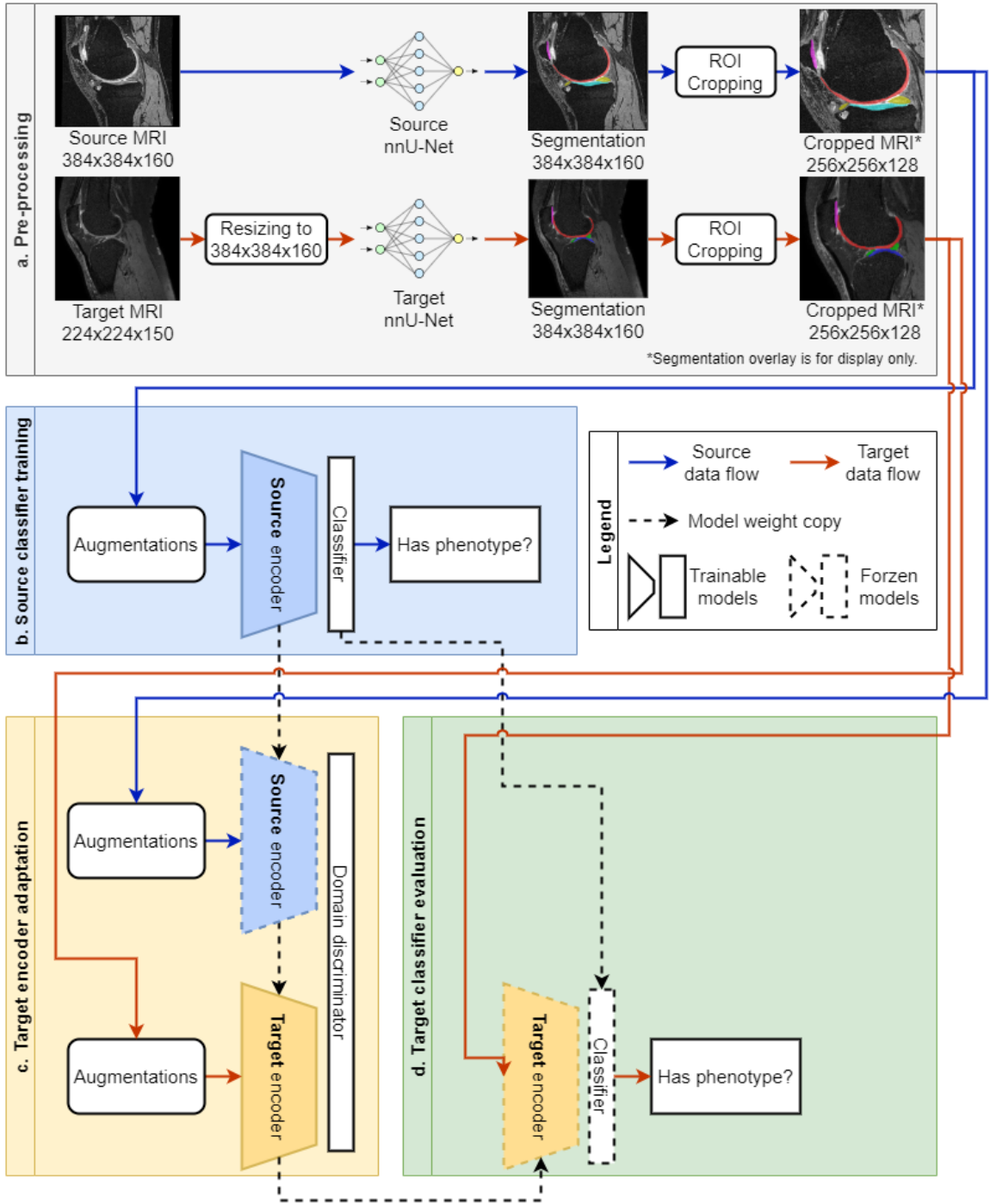

Figure 1. A systematic overview of the proposed UDA method. a. Magnetic resonance (MR) image pre-processing included image resizing, segmentation and ROI extraction. Source MR image: 3D DESS, target MR image: 3D TSE/FSE b. Source classifiers for each OA phenotype were trained using the source MR images and labels. c. The source encoders were adapted to the target data using UDA, forming target encoders for each OA phenotype. d. Target classifiers, consisting of the adapted target encoders and the classification heads trained during source encoder training, were evaluated on the target dataset. ROI, region of interest, UDA, unsupervised domain adaptation

## Unsupervised Domain Adaptation Pipeline

Figure 1 illustrates the proposed UDA pipeline adapted from the ADDA (11) framework. The four steps of the UDA pipeline are described below.

### MR Image Pre-processing

The target MR images were first resized to 384 × 384 × 160 to align with the source MR images (Figure 1a.). Two 2D nnU-Net (19) models were trained using the official implementation and default settings for providing segmentation masks. We trained the source and target segmentation models using 3D MR scans from the OAI-iMorphics dataset (141 for training, 35 for testing) and the labelled target dataset (17 for training, 8 for testing), respectively. Segmentation masks for source and target data samples were predicted by the nnU-Nets, which were used to locate the TFJ and PFJ. Using this location, 3D volumes (256 × 256 × 128) were cropped as the region of interest (ROI) for later use.

### Source Classifier Training

The source classifiers, which consist of a 3D DenseNet121 (20) encoder ('source encoder') and a classification head, were trained on the balanced source dataset (Figure 1b.) for each phenotype. The augmentation scheme involved adding random Gaussian noise, scaling the intensity by a randomly selected factor from [0.8, 1.2], rotating by a randomly chosen degree from [-10, 10] and scaling the image size by a randomly chosen factor from [1, 1.1]. Each augmentation had a probability of 50%. We trained the classifiers with a batch size of 2, focal loss (21) with a gamma of 1, an Adam (22) optimiser with a learning rate of $10^{-6}$ and a weight decay of $10^{-3}$. The training process stopped when the validation loss increased by three times, and the model with the best area under the precision-recall curve on the validation set was selected.

### Target Encoder Adaptation

As shown in Figure 1c., the domain adaptation process involved a source encoder with frozen weights, a target encoder initialised with the weights of the source encoder, and a randomly initialised domain discriminator. We adopted the hyperparameters from the study of Jiang et al. (8), in which the adaptation was optimised by a domain adversarial loss (23) and a stochastic gradient descent (24) optimiser with an initial learning rate of 0.001, a momentum of 0.9 and weight decay of $10^{-3}$ at a batch size of 2. The learning rate decays according to the following formula:

$$\alpha_{n+1} = \alpha_n \times (1 + \gamma \times n)^{-\lambda},$$

where $\alpha_{n+1}$ and $\alpha_n$ are the learning rates at the next epoch and the current epoch, respectively; $n$ is the current epoch count and $\gamma$ and $\lambda$ are hyperparameters set at 0.0003 and 0.75, respectively. The same image augmentations used for the source encoder training were applied to both the source and target MR images. Each training was run for 50 epochs, and the models from the last epoch were selected.

### Target Classifier Evaluation

We attached the trained classification head from the source encoder training to the adapted target encoder, forming the target classifier (Figure. 1d.). We trained a non-UDA classifier using the phenotype labels from the target dataset. The hyperparameters set for the source encoder training were used for the non-UDA classifier.

### Statistical Analysis

For each phenotype, the balanced source dataset was split into training (70%), validation (10%) and test (20%) sets to train the source encoder. The classifiers (with and without UDA) were trained on the target dataset by a leave-one-out strategy that used 49 samples for training and one sample for testing. During the target encoder adaptation, we continuously fed samples from the 3,116-sample source dataset while looping 49 target training samples for each epoch.

Dice similarity coefficient (DSC) scores were used to evaluate the segmentation models. The area under the receiver operating characteristic curve (AUROC), sensitivity, specificity and accuracy were used to assess the source and target classification models. The McNemar test was used to compare the performance of the classifiers trained with and without UDA. We considered $p < 0.05$ to indicate statistical significance.

### Implementation

The proposed system and scripts were implemented and tested using Python 3.10, PyTorch (25) 1.10, MONAI 0.8.1, PyTorch Lightning 1.6.3 and Transfer Learning Library (8) 0.2 (UDA only). An NVIDIA (Santa Clara, CA, USA) Quadro RTX5000 graphics processing unit was used to run the deep learning experiments. Statistical analyses were conducted using SPSS 28.0 (IBM, Armonk, NY, USA).

## Results

### Demographics

Table 2. Demographics and distribution of phenotypes among three datasets

| Characteristic | Source Dataset | Balanced Source Dataset | Target Dataset |
| --- | --- | --- | --- |
| Age (y) | 61.08 ± 8.97 | -- | 61.94 ± 11.57 |
| BMI (kg/m$^2$) | 28.56 ± 4.77 | -- | -- |
| Male (%) | 39.92 (1244/3116) | -- | 30 (15/50) |
| Female (%) | 60.08 (1872/3116) | -- | 70 (35/50) |
| Left knee (%) | 37.36 (1164/3116) | -- | 42 (21/50) |
| Right knee (%) | 62.64 (1952/3116) | -- | 58 (29/50) |
| Cartilage/meniscus (%) | 3.42 (106/3104) | 33.33 (106/318) | 8 (4/50) |
| Subchondral bone (%) | 10.29 (320/3110) | 33.33 (320/960) | 10 (5/50) |

Note. Age and body mass index (BMI) are presented as the mean ± standard deviation. BMI = body mass index.

The source dataset collected from the OAI dataset (13) includes 3,116 knee subjects with an average age of 61.08 years and an average BMI of 28.56 kg/m$^2$. In the target dataset, the average age of the participants was 61.94 years. Detailed demographics and data distribution for both datasets were reported in Table 2.

## MR Image Pre-processing

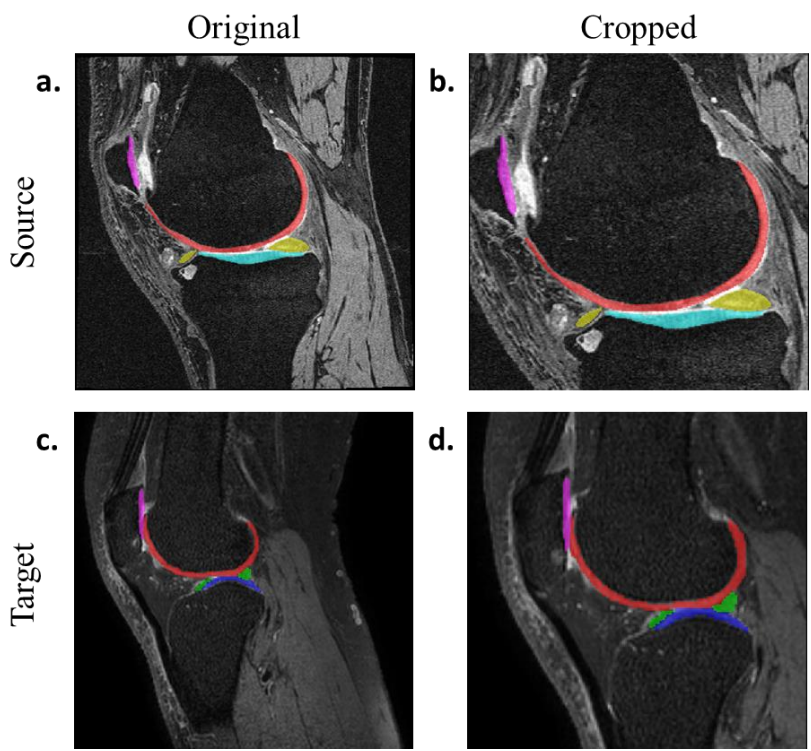

Figure 2. Examples of tissue segmentation. Columns 1 (a & c) and 2 (b & d) represent the original and cropped images with segmentation masks, respectively. Rows 1 (a & b) and 2 (c & d) are MR images from the source and target datasets, respectively.

Both segmentation models (nnU-Net) performed well, with DSC scores ranging from 0.81 to 0.92 (Table 3). We consider these scores to be satisfactory because the purpose of these predicted segmentation masks was to locate the PFJ and TFJ. An example of automatic ROI selection is shown in Figure 2. The ROI covered the TFJ and PFJ.

Table 3. Dice similarity coefficient scores of the source and target nnU-Net

| Knee compartment | Femoral Cartilage | Lateral Meniscus | Lateral Tibial Cartilage | Medial Meniscus | Medial Tibial Cartilage | Patellar Cartilage |
|---|---|---|---|---|---|---|
| Source nnU-Net | 0.92 ± 0.02 | 0.91 ± 0.02 | 0.92 ± 0.02 | 0.87 ± 0.05 | 0.90 ± 0.03 | 0.87 ± 0.08 |
| Target nnU-Net | 0.85 ± 0.02 | 0.85 ± 0.08 | 0.81 ± 0.05 | 0.85 ± 0.12 | 0.81 ± 0.06 | 0.81 ± 0.10 |

Note. Numbers are the mean ± standard deviation from the leave-out test set.

## Performance of Source Classifier

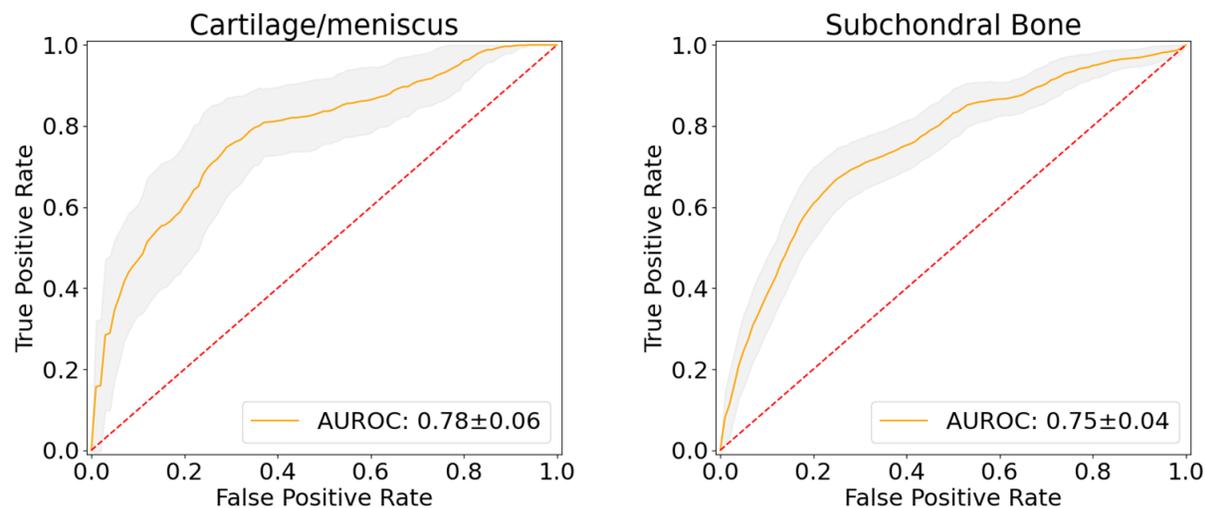

Figure 3. ROC and AUROC for the source encoders in the phenotype classification task. The orange curves are the averages of the 100 bootstrapping events. The shadows indicate the range of the curves. The dashed line indicates an AUC of 0.5. AUROC scores are shown as the mean ± standard deviation. ROC, receiver operating characteristics; AUROC, area under the receiver operating characteristic curve

The source classifiers for knee OA phenotype classification were evaluated on the hold-out test set by bootstrapping 100 times. Figure 3 shows the ROC for the classifiers. For the cartilage/meniscus and subchondral bone phenotype classifications, the means ± standard deviations of AUROC scores were 0.78 ± 0.06 and 0.75 ± 0.04, the sensitivities were 0.44 ± 0.11 and 0.64 ± 0.07, the specificities were 0.92 ± 0.04 and 0.76 ± 0.04, and the accuracies were 0.76 ± 0.05 and 0.72 ± 0.03, respectively.

## Performance of Target Classifiers

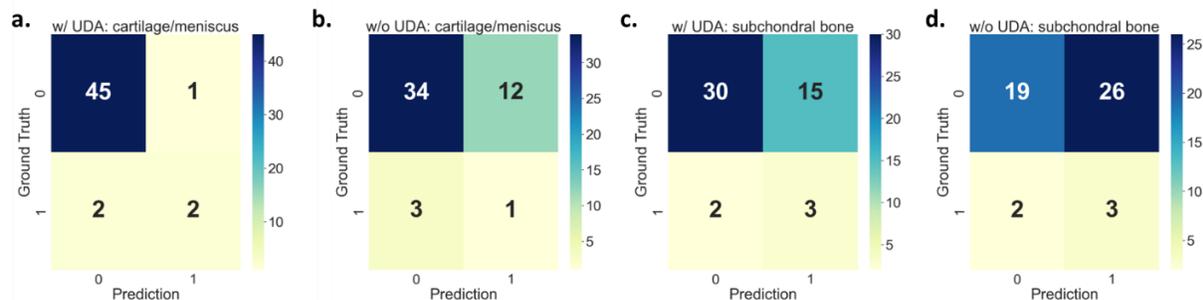

Figure 4. Confusion matrix of the classifiers trained with (a & c) and without UDA (b & d). w/, with; w/o, without; UDA, unsupervised domain adaptation

The classification metrics are shown in Table 4, and the corresponding confusion matrix is provided in Figure 4. UDA training improved the performance of the target classifiers compared with the non-UDA classifiers. Improvements in all parameters (AUROC, sensitivity, specificity and accuracy) were achieved with the use of UDA. Compared with the classifiers trained without UDA, the UDA-trained classifiers performed significantly better at classifying the cartilage/meniscus (McNemar test, $p = 0.013$) and subchondral bone (McNemar test, $p = 0.027$) phenotypes.

Table 4. Performance of phenotype classifiers evaluated on the target dataset

Phenotype: cartilage/meniscus

| Classifier | AUROC | Sensitivity (%) | Specificity (%) | Accuracy (%) |
|---|---|---|---|---|
| With UDA | 0.83 | 50 (2/4) | 97.83 (45/46) | 94 (47/50) |
| Without UDA | 0.52 | 25 (1/4) | 73.91 (34/46) | 70 (35/50) |

Phenotype: subchondral bone

| Classifier | AUROC | Sensitivity (%) | Specificity (%) | Accuracy (%) |
|---|---|---|---|---|
| With UDA | 0.67 | 60 (3/5) | 66.67 (30/45) | 66 (33/50) |
| Without UDA | 0.53 | 60 (3/5) | 42.22 (19/45) | 44 (22/50) |

Note. AUROC = area under the receiver operating characteristic curve, UDA = unsupervised domain adaptation

## Discussion

In this study, we proposed a UDA application for MRI-based OA phenotype classification. The proposed method adapted the CNN encoder trained on the large, publicly available OAI dataset to a smaller, locally collected target dataset. The performance of the target classifier trained with UDA, as evaluated by the AUROC, sensitivity, specificity and accuracy, improved significantly compared with the classifier trained without UDA. The target classifier trained with UDA successfully captured the crucial information from the source dataset for knee OA phenotype classification and adapted it to the target dataset without target data labels, despite the differences in MRI sequences, scanner vendors, acquisition parameters and patient groups between the two datasets.

UDA is widely used in various medical imaging tasks, including segmentation (26–28) and diagnostics (29), and has been successfully applied on images of various organs, including the heart (28), brain (30), breast (31), liver (26) and knee (27). UDA applications in medical imaging address the issue of insufficient data in the target dataset. To the best of our knowledge, applying UDA to OA phenotype classification has not been explored previously and the method developed in this study fills that gap. Our proposed method may accelerate downstream research on diagnostics and prognostics in local patient groups using deep learning and address the challenges of limited data and high labelling costs that hinder such research.

Our work contributes to downstream research by developing a system based on clinically validated ground truth and standard TSE/FSE MRI acquisition. The clinically validated phenotypes were proposed by Roemer et al. (18), who conducted a case–control association analysis of phenotypes with multiple clinical assessments for 475 knee subjects from the Foundation for National Institutes of Health OA Biomarkers Consortium cohort, a subset of the OAI dataset (32). The analysis revealed a relationship between the subchondral bone phenotype and the risk of radiographic OA progression. In a longitudinal analysis of the entire OAI dataset by Namiri et al. (22), correlations between all phenotypes and concurrent structural OA were discovered. 2D TSE/FSE sequences are widely used in clinical routines and research projects, such as the OAI (13), and 3D variants have been used for deep-learning-based OA studies. For example, Astuto et al. (3) presented a knee OA staging system with multiple binary classifiers from 1,786 3D TSE/FSE MR images, yielding AUROC values ranging from 0.83 to 0.93 for all knee compartments. Namiri et al. (4) performed severity staging on anterior cruciate ligament injuries using a 3D TSE/FSE MR image dataset containing images from 1,243 knee subjects, yielding 89% and 92% accuracy with 3D and 2D CNNs, respectively. These previous studies have generated promising results using supervised training with large high-quality datasets. Our work extends the successful use of TSE/FSE knee MR images to the UDA approach and enables downstream clinical studies using smaller datasets.

Although UDA improved the classification performance compared with the non-UDA variant, we were limited by small sample sizes and an imbalanced data distribution. The small sample size of the target dataset ($n = 50$) constrained the statistical power and deep-learning performance. Additionally, the sample size and spatial geometry imbalance of our dataset affected the classification power. The patient groups had much smaller sample sizes than the control groups, and the geometry size of the image features that defined the phenotypes was small compared with the geometry size of the 3D knee volume. The small sample size and geometry imbalance may have led to poor performance of the CNN. We attempted to overcome the sample size inequality using focal loss (21) and the geometry imbalance using ROI cropping. Future investigations to address the data imbalance issue are of interest. For example, zero-shot learning (33,34), which learns from samples of the majority classes to improve the minor-class classification, may be applied. The pathology detection approach proposed by Desai et al. (35) has the potential to address the geometry imbalance issue by converting the phenotype classification to phenotype-related feature detection tasks.

In conclusion, we report a UDA approach for automatically classifying knee OA phenotypes in 3D TSE/FSE MR images. The proposed UDA implementation transfers task-specific information from the publicly available large OAI dataset to a small in-house dataset. Our application of UDA significantly improved the phenotype classification performance, making downstream clinical research on local patient groups possible. As UDA is transferable to arbitrary medical imaging tasks, it may provide a springboard for further research across institutions, imaging devices, acquisitions and patient groups.

## Acknowledgements

We would like to acknowledge Ben Chi Yin Choi and Cherry Cheuk Nam Cheng for assistance in patient recruitment and MRI exams. This study was supported by a grant from the Innovation and Technology Commission of the Hong Kong SAR (Project MRP/001/18X), and a grant from the Faculty Innovation Award, the Chinese University of Hong Kong.